\documentclass[10pt,journal,compsoc]{IEEEtran}

\usepackage{graphicx}
\usepackage{dcolumn}
\usepackage{bm}
\usepackage{amsmath}
\usepackage{amssymb}
\ifCLASSOPTIONcompsoc
\usepackage{tabularx,booktabs}
\usepackage{algorithm}
\usepackage{algpseudocode}

  \usepackage[nocompress]{cite}
\else
  \usepackage{cite}
\fi

\ifCLASSINFOpdf
\else
\fi

\begin{document}

\title{A principled approach for weighted multilayer network aggregation}

\author{Junyao~Kuang, Caterina~Scoglio
\IEEEcompsocitemizethanks{\IEEEcompsocthanksitem Junyao Kuang is with the Department of Electrical and Computer Engineering, Kansas State University, Manhattan, KS, 66502. E-mail: kuang@ksu.edu 
\IEEEcompsocthanksitem Caterina Scoglio is with the Department of Electrical and Computer Engineering, Kansas State University, Manhattan, KS, 66502. E-mail: caterina@ksu.edu}
\thanks{}}


\IEEEtitleabstractindextext{%
\begin{abstract}
A multilayer network depicts different types of interactions among the same set of nodes. For example, protease networks consist of five to seven layers, where different layers represent distinct types of experimentally confirmed molecule interactions among proteins. In a multilayer protease network, the co-expression layer is obtained through the meta-analysis of transcriptomic data from various sources and platforms. While in some researches the co-expression layer is in turn represented as a multilayered network, a fundamental problem is how to obtain a single-layer network from the corresponding multilayered network. This process is called multilayer network aggregation. In this work, we propose a \emph{maximum a posteriori} estimation-based algorithm for multilayer network aggregation. The method allows to aggregate a weighted multilayer network while conserving the core information of the layers. We evaluate the method through an unweighted friendship network and a multilayer gene co-expression network. We compare the aggregated gene co-expression network with a network obtained from conflated datasets and a network obtained from averaged weights. The Von Neumann entropy is adopted to compare the mixedness of the three networks, and, together with other network measurements, shows the effectiveness of the proposes method. 

\end{abstract}

}
\maketitle

\IEEEdisplaynontitleabstractindextext
\IEEEpeerreviewmaketitle

\IEEEraisesectionheading{\section{Introduction}\label{sec:introduction}}

\IEEEPARstart{N}{etworks} 
are widely used to describe the interactions between nodes within a system. The interactions can be either represented by unweighted edges or quantified by weights on edges. In terms of epidemic analysis, weighted networks provide more accurate results than their unweighted counterparts according to \cite{gemf1, gemf2}. In different types of networks, edges are detected in various ways. For example, in a social network, an edge is detected if two people recognize each other or meet up at the same places during a period \cite{newman1, mit}. A weighted social network can be obtained by simply asking people how often they meet their friends during a period \cite{fs1, fs2}. Similarly, in constructing gene co-expression networks, the weights of edges are determined through correlation coefficients, and an unweighted gene co-expression network can be obtained by selecting edges with correlation coefficients greater than a threshold \cite{pcc1}. In papers \cite{pcc1,pcc2,pcc3}, the Pearson correlation is used to determine edges for gene co-expression networks. The gene expression levels are recorded after subjects receive different stimuli, the Pearson correlation coefficients are calculated for every pair of genes and a fixed threshold is used to select edges. For example, if two genes have a Pearson correlation coefficient greater than 0.8, then an edge is detected between the two genes \cite{pcc3}.

However, when the data come from various sources and platforms, it is difficult to conflate different datasets and find a single-layer network. For example, to construct a gene co-expression network, authors in \cite{pcc1} construct a network from a conflated dataset, where gene expression values are tested under different platforms, such as RNA-sequencing and microarray. The gene expression values are normalized through a median-shifted method, and the Pearson correlation coefficient is used to detect edges. The drawback of the method is that each expression value must be recorded more than once in each condition. The correlation coefficients in the co-expression map are significantly affected by the normalization method, and most of the correlation coefficients are greater than 0.9, which 
means that most of the genes in the network are correlated. To avoid mixing different experiments together, the multilayer network is used to depict different types of experiments in some circumstances \cite{fs2, pcc2}. For examples, authors in \cite{pcc2,pcc3,multi} use multilayer networks to represent the relations of genes. However, when there are too many layers, it is difficult to analyze the network properties, such as centralities and degree distribution \cite{simp,simp2,simp3}. Besides, in the co-expression network, the layers obtained through different types of experiments need to be aggregated as a single layer in a more complex multilayer network \cite{reduce}. In this work, we are interested in constructing a single layer network by aggregating a multilayer network which is obtained from various sources.

Although various methods have been used to construct single layer network, but the methods are either too difficult to implement or based on random criteria. Paper \cite{aggdirect} compares three methods for multilayer network aggregation and link prediction, including a direct aggregation, a result aggregation, and a layer co-analysis method. In all of the three methods, the networks are mapped to vectors by training the \emph{node2vec} neural network \cite{node2vec}. The drawbacks of the algorithm are obvious; it is necessary to have a large dataset to train the neural network, and a high-performance processor is required. Authors in paper \cite{reduce} use the Von Neumann entropy to evaluate the reducibility of a multilayer network. And the authors adopt a greedy agglomerative hierarchical clustering algorithm to explore the space of partitions of different layers, but the layers are aggregated directly. In paper \cite{simp,simp2}, the authors introduce a selection method, which selects nodes or edges of interest with high centralities to obtain the final single-layer network. Also, a method that filters nodes and edges with lower centralities is introduced to obtain a single-layer network. Paper \cite{pcc2} presents a method that constructs a network for each dataset, and an edge is confirmed in if it is observed in more than two datasets. To obtain a single layer gene co-expression network, a rank aggregation method is introduced in \cite{pcc3}. In paper \cite{invest}, the authors use a statistical method for aggregating unweighted networks, in which the probability of observing an edge is decided by the significance level obtained by Bonferroni correction. 

In this work, we propose a simple and principled method to aggregate multilayer networks, which is the main contribution. The method introduced in this paper is based on statistical distribution of edge weights across the layers. Inspired by \cite{newman1,newman2,newman3,newman4}, we adopt the \emph{maximum a posteriori} (MAP) estimation and expectation-maximization algorithm to determine the weights of edges in the target single-layer network. In the derivation of the proposed method, we make the following assumptions. First, each layer could contain part or all of the nodes of interest, and the edge weights of different layers are non-negative. Second, we assume that the different edges in the multilayer network are independent and identically distributed. Also, the expected weight of each edge follows Poisson distribution. Further, the ensemble distribution of all the weights of edges is assumed to follow the exponential distribution, as introduced in \cite{exp}.

The paper is organized as follows. In section two, we describe the model for constructing the target layer with the MAP estimation. In Section three, to validate the effectiveness of the proposed method, we implement the algorithm to aggregate an \emph{Anopheles Gambiae} gene co-expression network and a friendship multilayer network. For the co-expression network, we compare the properties of the resulted network with a network constructed from conflated datasets and a network aggregated directly, and it shows that the proposed method can reduce experimental uncertainty. We conclude the article in section four and discuss future works.

\section{MAP Estimation-Based Network Aggregation}\label{sec2}
We denote the target network as $A$, and assume there are $N$ nodes across all of the $D$ layers. Note that different layers could have different sets of nodes. For every pair nodes $i$ and $j$ in the network $A$, we assume the expected weight is $\lambda_{ij}$. Therefore, the problem can be simplified to determining the parameter set $\lambda=\{\lambda_{ij}\ |\ i\leq N,j\leq N\}$, given the parameters of multilayer network $D$. In this method, we use a probabilistic model, the likelihood of the model parameters given $D$ is $P(\lambda\ |\ D)$. Applying the Bayes theorem, we have
\begin{align}
\label{eq:1}
    P(\lambda\ |\ D)=\frac{P(D\ |\ \lambda) P(\lambda)}{P(D)},
\end{align}
where $P(D\ |\ \lambda)$, $P(\lambda)$ and $P(D)$ are the likelihood of the data sets $D$ under network constructed through parameter set $\lambda$, the prior probability, and the marginal likelihood, respectively. 
In a particular dataset, it is possible that only a part of the $N$ nodes are tested, so we can only obtain the weights for some of the edges. To determine the expression of $P(D\ |\ \lambda)$, we assume the relationship between $i$ and $j$ is detected $k_{ij}$ times in the $D$ layers, and the weight is denoted as $\xi_{ij}^\alpha$ in a particular layer $\alpha$. For the node pair $i$ and $j$, we assume the weight of the edge follows a Poisson distribution, so the probability of $\xi_{ij}^\alpha$ can be expressed as
\begin{align}
\label{eq:2}
    P(\xi_{ij}^\alpha\ |\ \lambda_{ij})=\frac{e^{-\lambda_{ij}} \lambda_{ij}^{\xi_{ij}^\alpha}}{\xi_{ij}^\alpha!}.
\end{align}

Further, the likelihood function of $\lambda_{ij}$ is then 
\begin{align}
\label{eq:3}
    L(\lambda_{ij}\ |\ \xi_{ij}^1,...,\xi_{ij}^{k_{ij}})=&
    \prod_{\alpha=1}^{k_{ij}} \frac{e^{-\lambda_{ij}} \lambda_{ij}^{\xi_{ij}^\alpha}}{\xi_{ij}^\alpha!}\nonumber\\
    =& \frac{e^{-k_{ij} \lambda_{ij}} \lambda_{ij}^{\sum_\alpha^{k_{ij}} \xi_{ij}^\alpha}}{\prod_{\alpha=1}^{k_{ij}} \xi_{ij}^\alpha!}.
\end{align}

In maximum likelihood estimation (MLE), the prior probability $P(\lambda)$ is regarded as a constant, and the prior probability will be leave out when applying the expectation-maximization algorithm to find the optimal parameters. On the contrary, the prior probability is a requirement in MAP estimation. The prior probabilities are generally provided by pilot studies \cite{multi}. However, the prior probabilities for each pair of nodes are unknown in our assumptions. Instead, we assume a prior distribution for the parameter set $\lambda$. In the target network, we assume the edge weights follow exponential distribution \cite{exp}, which is given by
\begin{align}
\label{eq:4}
    P(\lambda_{ij})=\theta e^{-\lambda_{ij} \theta},
\end{align}
where $P(\lambda_{ij})$ and $\theta$ are the prior probability of $\lambda_{ij}$ and the rate parameter of the exponential distribution, respectively. The posterior probability of the target network is then 
\begin{align}
\label{eq:5}
    P(\lambda\ |\ D)=\frac{1}{P(D)}\prod_{i,j}\frac{e^{-k_{ij} \lambda_{ij}} \lambda_{ij}^{\sum_\alpha^{k_{ij}} \xi_{ij}^\alpha} \theta e^{-\lambda_{ij} \theta}}{\prod_{\alpha=1}^{k_{ij}} \xi_{ij}^\alpha!}.
\end{align}

However, equation \ref{eq:5} is still intractable. To maximize the posterior probability, we take the logarithm form of the expression. Leaving out constant term $P(D)$ and the factorial terms. We have
\begin{align}
\label{eq:6}
    L(\lambda\ |\ D)=&\sum_{i,j}[-k_{ij}\lambda_{ij}
    +\sum_\alpha^{k_{ij}} \xi_{ij}^\alpha log(\lambda_{ij}) \nonumber\\
    +& log(\theta) -\lambda_{ij}\theta].
\end{align}

We use the expectation-maximization algorithm to find the optimal values of $\{\lambda_{ij}\}$. Taking the derivation of equation \ref{eq:6} in terms of $\lambda_{ij}$ and $\theta$, the estimates for $\lambda_{ij}$ and $\theta$ are
\begin{align}
\label{eq:7}
\theta=\frac{N^2}{\sum_{i,j} \lambda_{ij}},
\end{align}
\begin{align}
\label{eq:8}
\lambda_{ij}=\frac{\sum_\alpha^{k_{ij}} \xi_{ij}^\alpha}{k_{ij}+\theta}.
\end{align}

The optimal $\lambda_{ij}$ is calculated by updating $\theta$ and $\lambda_{ij}$ alternatively. In our experiments, random initial values are assigned to the two parameters. 

Since the edge weight is assumed to follow Poisson distribution, so the expected edge weight between node $i$ and node $j$ in the target network is exactly $\lambda_{ij}$. The distribution of $\{ \lambda_{ij} \}$ in the target network is an exponential distribution with rate parameter $\theta$.

In aggregating multilayer networks, it is difficult to evaluate the target network, since there is no ground truth network for comparison. To intuitively compare how mixedness of different networks, we adopt the Von Neumann entropy itntroduced in \cite{reduce}. In quantum mechanics, a system can be described by a semidefinite positive matrix with eigenvalues summing up to one, and the Von Neumann entropy is used to evaluate the mixedness of the system. A larger Von Neumann entropy represents a more mixed state. According to \cite{reduce}, the statistical ensemble of a network can be measured by the Von Neumann entropy. To evaluate the distinguishability between the multilayer networks and the aggregated network, the authors use Von Neumann entropy to study the reducibility of multilayer networks in \cite{reduce}.

The Laplacian matrix \emph{L} of a graph has non-negative eigenvalues. Therefore, the Laplacian matrix can be used to measure the Von Neumann entropy. The Von Neumann entropy is defined as 
\begin{align}
\label{eq:9}
h_A=-\sum_{i=1}^{N}\gamma_i log(\gamma_i).
\end{align}
where $\gamma_i$ is the eigenvalue of the Laplacian matrix. The elements in the Laplacian matrix are normalized by twice the sum of edge weights. In this work, we use the Von Neumann entropy to evaluate the mixedness of different single-layer networks.
\section{Numerical Results}\label{sec3}
It is difficult to validate the proposed method in section \ref{sec2}, since there is no ground truth network for comparison \cite{newman1}. Firstly, we aggregate a single layer gene co-expression network and compare the properties of the network with two single-layer networks obtained through traditional methods. Secondly, we verify that the method does not distinguish the layers when aggregating unweighted multilayer networks, which is the expected result.

\subsection{Results on weighted multilayer network}
In the first case, we construct a single layer gene co-expression network through the MAP method introduced in section \ref{sec2} (for short, we call it MAPNet), and the nodes in the network are genes from \emph{Anopheles Gambiae}. To compare the results, we construct a network (ConNet) from the conflated datasets, and a network (AveNet) with edge weights averaged across all of the layers.

\subsubsection{Data pre-process}
In detecting gene co-expression networks, the genes are tested under different conditions, and the experiments are conducted under various platforms \cite{koutsos,mari,cassone1,goltsev,mendes,cassone2,baker,rund}. For example, to study how the genes functioning in different development stages, some researchers \cite{goltsev, koutsos} measure the expression levels of \emph{Anopheles Gambiae} genes in different embryonic and developmental stages. Also, some researchers \cite{mari, mendes} study how blood meal and plasmodium falciparum infections affect the expression levels of genes, etc. In this work, we apply the construction method to 290 genes of interest which come from 11 experiments (or datasets). Each dataset contains 12 to 20 conditions, and there are 156 conditions in total. However, in each dataset, some genes could have missing values or even not tested. For example, the two nodes $x$ and $y$ have the following expression values:

$x=[\times,o,\times,\times,o,\times,\times,\times,\times,\times,\times,\times,\times,\times,\times,o,\times]$

$y=[\times,\times,\times,\times,o,\times,\times,\times,\times,\times,\times,\times,\times,o,\times,o,o]$.

Where there are 17 conditions in the dataset. $"\times"$ represents expression value, and $"o"$ represents a missing value, the number of paired elements between nodes $x$ and $y$ is 12. Therefore, the edge weight between the two nodes can be calculated through the Pearson correlation. Note that only gene pairs with at least five paired elements are considered, and negative coefficients are set to zeros. So the range of edge weights is in range [0, 1]. In our case, we normalize the edge weights to integers in the range [0, 10]. Therefore, we can obtain an eleven-layer weighted network. As a comparison, we construct a single-layer network, i.e., ConNet, from the 156 conflated conditions. Since the 156 conditions come from 11 datasets, we normalize the expression values of each condition by the so-called z-score normalization, i.e., $z_i=\frac{x_i-\bar{x}}{\sigma}$, where $\sigma$ is the standard deviation of all the expression values in the condition. Similarly, we compute the Pearson correlation coefficients of all the gene pairs with at least five paired elements and set negative values to zeros.

\subsubsection{Network aggregation and comparison}

\begin{figure}[h]
\includegraphics[width=\linewidth]{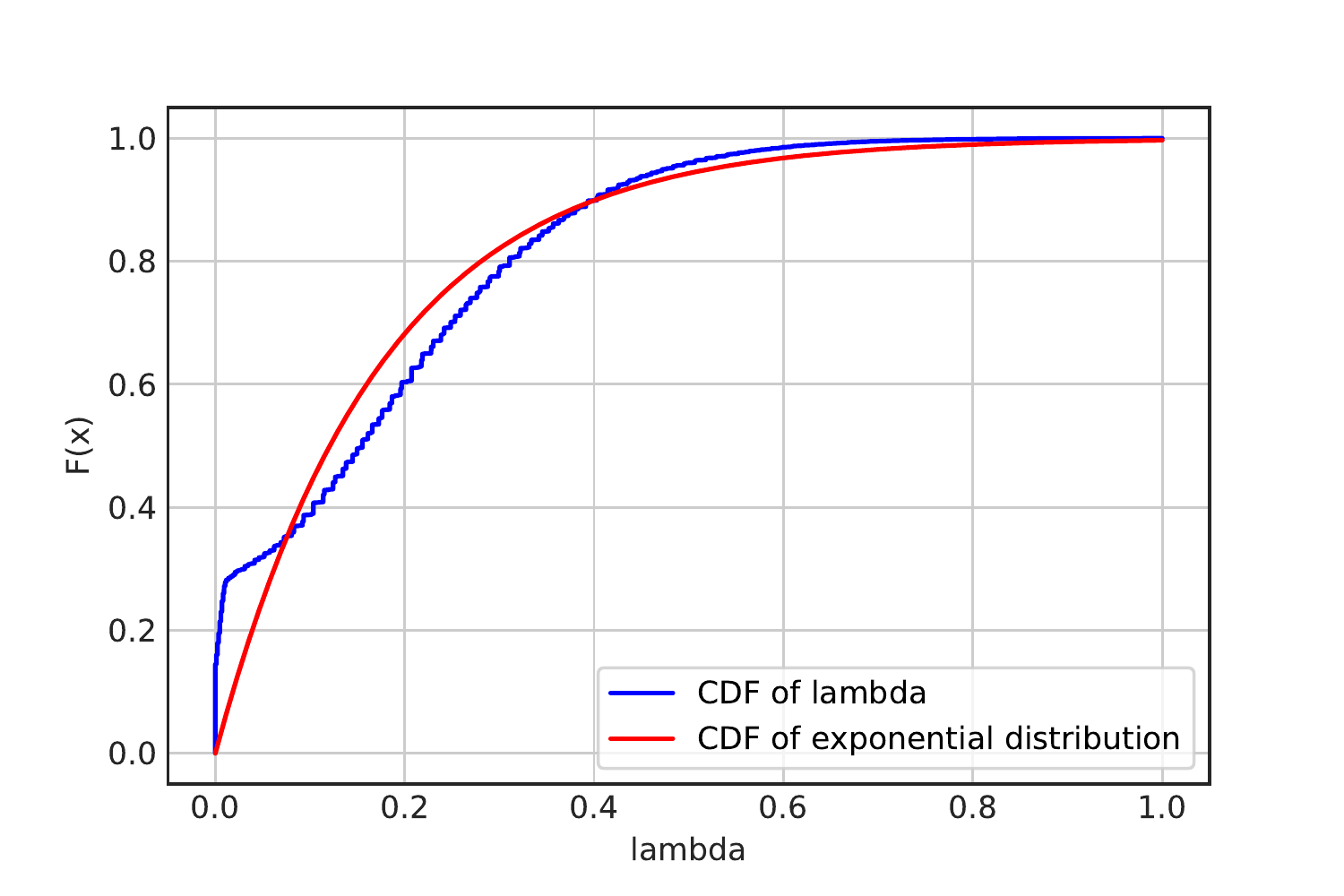}
 \caption{CDF of the edge weights distribution and the corresponding exponential distribution.}
 \label{fig:fig1}
\end{figure}

We aggregate the multilayer network through the MAP method introduced in section \ref{sec2}. In Fig. \ref{fig:fig1}, we show the cumulative distribution function (CDF) of edge weights in the aggregated network, as well as the corresponding exponential distribution. the coefficient of determination is 0.93 which means that the exponential distribution fits well the edge weights distribution.

\begin{figure}[h]
\includegraphics[width=\linewidth]{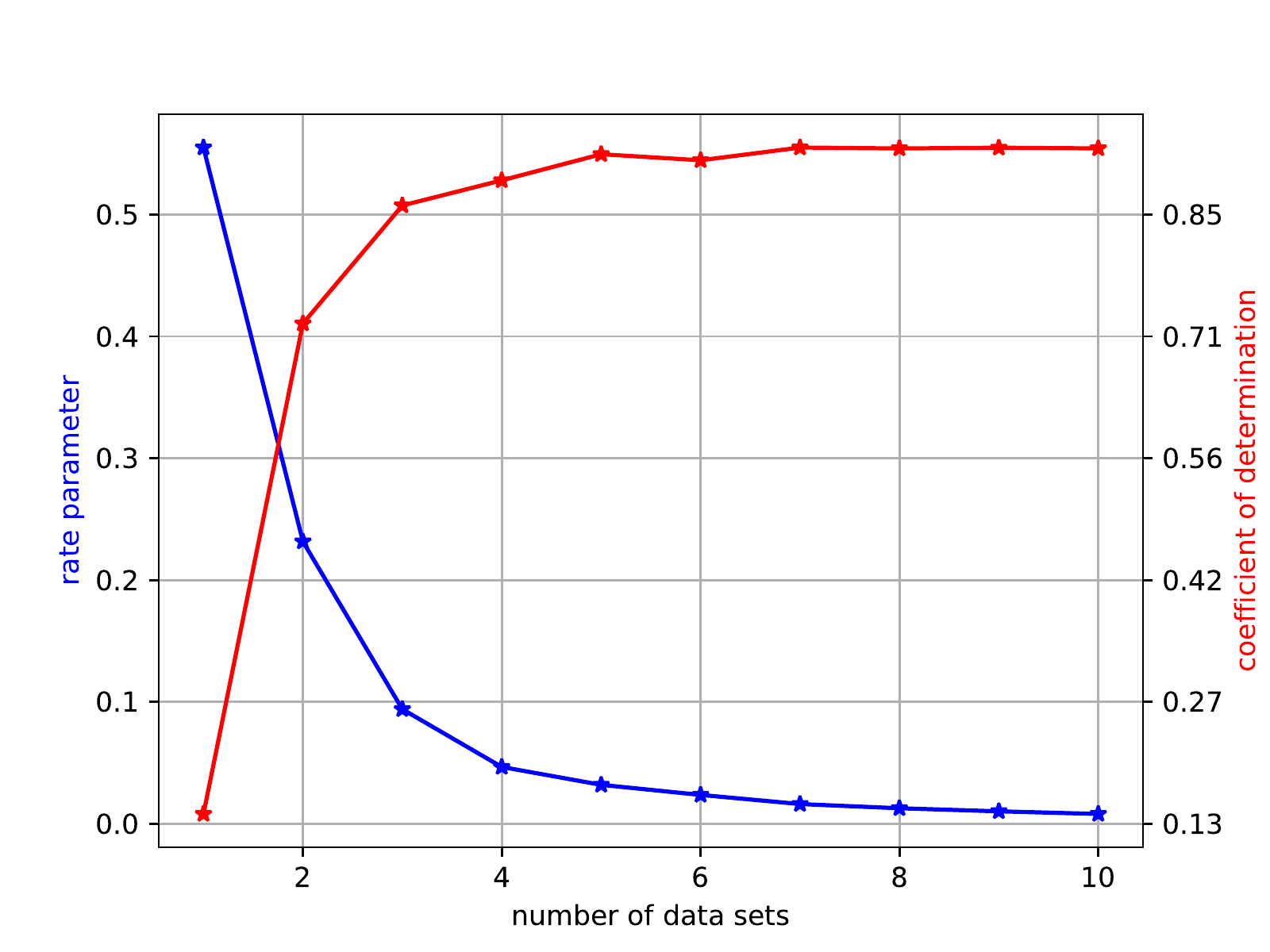}
 \caption{The relation between the number of datasets and rate parameters and coefficients of determination. Results are averaged over 10 combinations of the 11 datasets.}
 \label{fig:fig2}
\end{figure}

In Fig. \ref{fig:fig2}, we show how the number of layers affects the model. The red curve shows the coefficients of determination between the obtained exponential distribution and the distribution of the estimated $\{\lambda_{ij}\}$. As we can see, the coefficients of determination are above 0.9 when there are at least four datasets used, which means the exponential distribution fits well the edge weights distribution. The blue curve shows the variation of the rate parameters with respect to the number of datasets. The rate parameters converge when there are at least four datasets. It is because when there is more evidence, i.e., more layers, the edge weights are confirmed by more experimental results. The estimated edge weights are not affected by systemic errors. A network aggregated with 1 to 3 layers is highly affected by some unpredictable factors, such as experimental biases. The single layer network aggregated from more layers can help to improve the reliability of the constructed network and reduce uncertainty.

\begin{figure}[h]
\includegraphics[width=\linewidth]{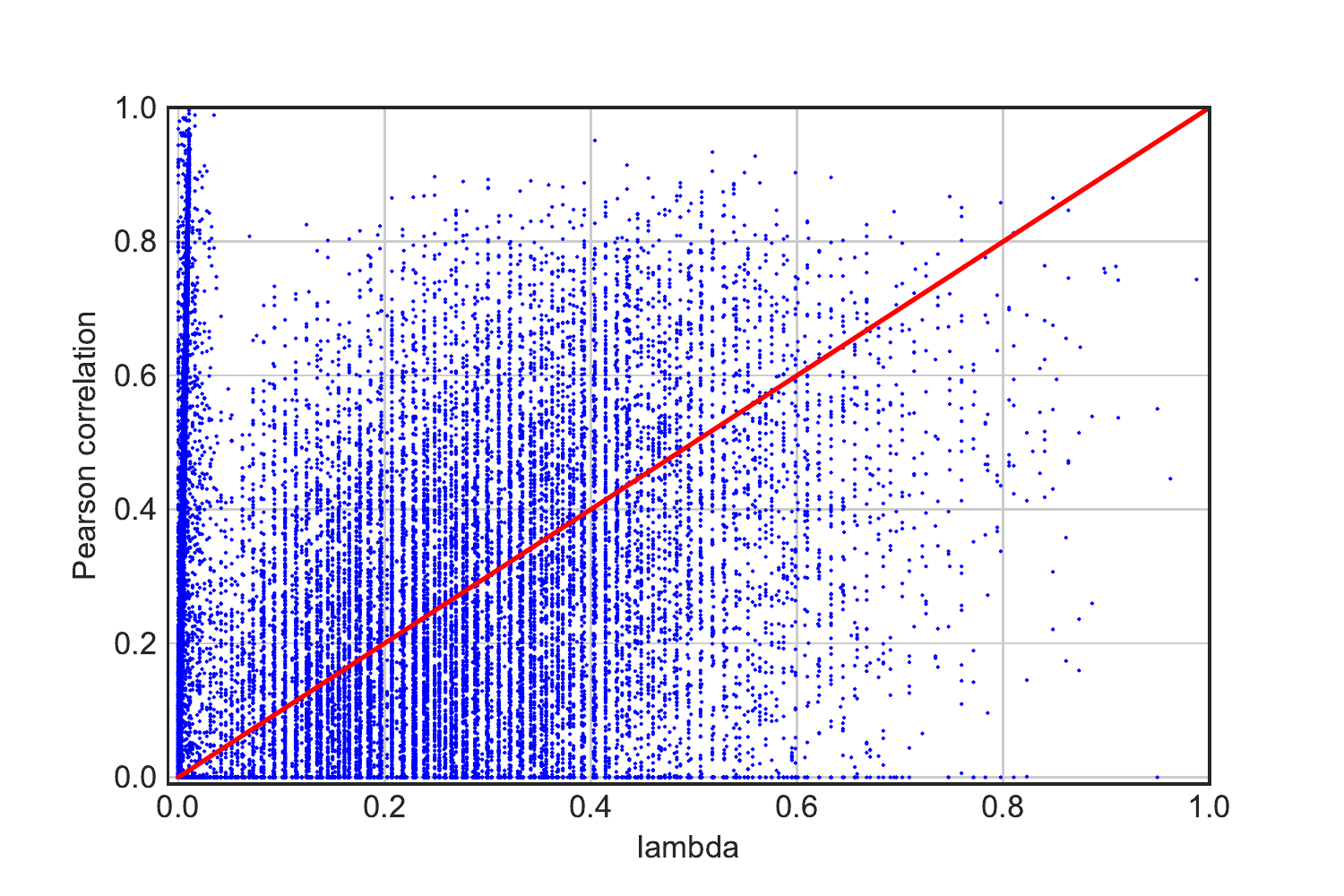}
 \caption{Comparison between Pearson correlation of conflated data set and $\lambda_{ij}$. The $\lambda_{ij}$ are obtained with 11 datasets.}
 \label{fig:fig3}
\end{figure}

\begin{figure}[h]
\includegraphics[width=\linewidth]{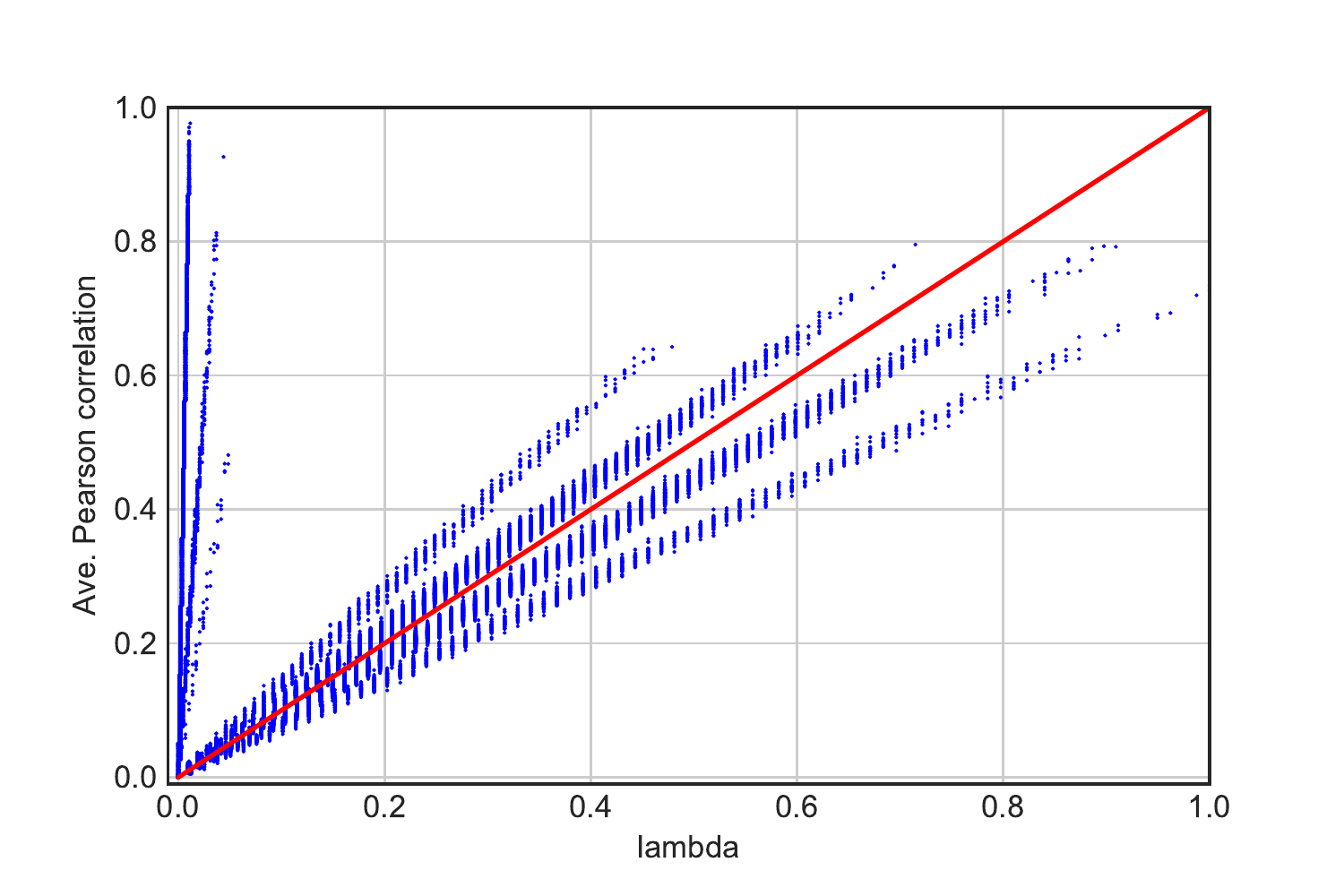}
 \caption{Comparison between averaged Pearson correlation coefficients and $\lambda_{ij}$. The $\{ \lambda_{ij} \}$ are obtained with 11 datasets.}
 \label{fig:fig4}
\end{figure}

The comparison between the ConNet and MAPNet is shown in Fig. \ref{fig:fig3}. The edge weights of the two networks are normalized by their corresponding maximums. As shown in the figure, the edge weights obtained through the two methods are heterogeneous. In the neighborhood of the vertical axis, there some edges have high correlation coefficients in the ConNet but low $\lambda_{ij}$ values in MAPNet. This is because, in some datasets, there are some genes with less than five paired elements. But after conflating the conditions of different datasets together, more genes are detected with at least five paired elements. More importantly, genes from the same dataset tend to have similar systemic biases, such as experimental manipulation and platform differences. This is the reason that some researchers aggregate multilayer networks to obtain a single-layer co-expression network\cite{pcc2, pcc3} rather than constructing a network from conflated datasets. 

\begin{figure}[h]
\includegraphics[width=\linewidth]{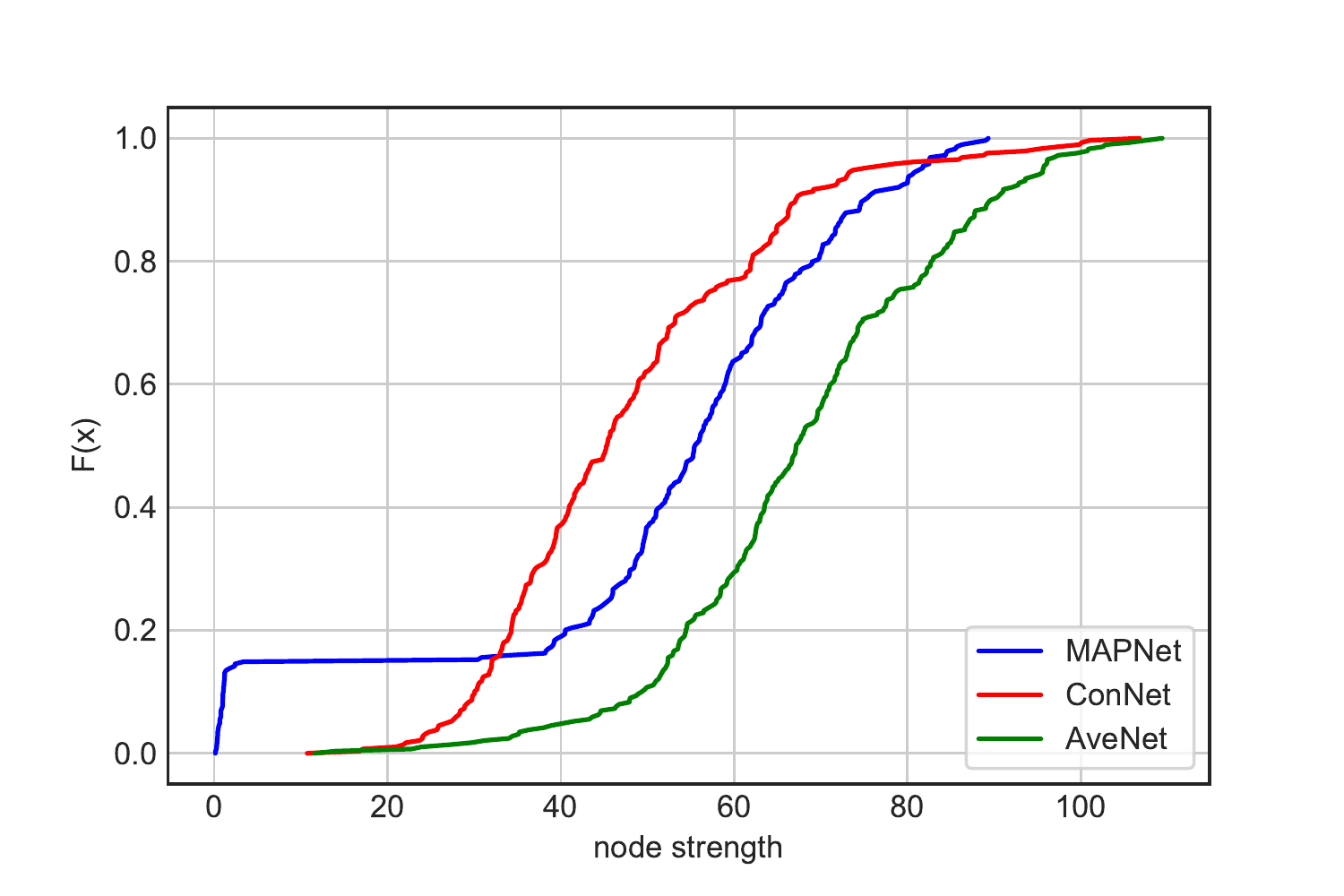}
 \caption{Edge weights of the aggregated network and the corresponding exponential distribution.}
 \label{fig:fig5}
\end{figure}

We then compare the $\{\lambda_{ij}\}$ with the averaged Pearson correlation coefficients, i.e., the AveNet. In Fig. \ref{fig:fig4}, it can be seen that there are also some values close to the vertical axis. The phenomenon is due to edges with low rate of presence are given lower edge weights. Also, we can see that the points are situated in different radial lines since edges detected in more datasets are given higher weights in the proposed method. Finally, we compare the node strength distribution of the three networks. The CDFs of the node strength distributions are shown in Fig. \ref{fig:fig5}. It can be seen, in the MAPNet, there is a significant number of nodes having low node strength, and this is accord with the results in Fig. \ref{fig:fig3} and Fig. \ref{fig:fig4}. 

\begin{table}[h]
\centering
\caption{\label{tab:tab1}%
Properties of the three networks
}
\begin{tabularx}{0.42\textwidth}{lccc}
\hline \hline
Properties & MAPNet & ConNet & AveNet \\
\hline
No. of nodes & 290 & 290 &290\\
Ave. node strength &  50.6 & 47.7 &68.0\\
Min. node strength & 0.1 &10.7 & 11.6\\
Max. node strength &  89.3 &106.8 & 109.4\\
Ave. clustering coef. & 0.16 &0.18 & 0.05\\
Eigenvector centrality & 0.052 & 0.053 &  0.056\\
Entropy & 9.6 & 9.8 &  10.1\\
\hline \hline
\end{tabularx}
\end{table}

Without setting thresholds for edges, we calculate the properties of the three weighted networks. The statistics of the three networks are shown in table \ref{tab:tab1}. As we can see, the MAPNet has a much smaller minimum node strength and maximum node strength. The proposed method can significantly filter out edges with low rate of presence across all of the layers. The ConNet and AveNet have higher entropy since the two methods tend to conserve all the information of the network as long as the edges are present in any of the datasets. On the contrary, the proposed method tends to conserve the edges that have a high rate of presence, and gives these edges higher weights as confirmed in Fig. \ref{fig:fig3} and Fig. \ref{fig:fig4}.

\subsection{Verification on unweighted multilayer network}
In the second case, we implement the proposed method to an unweighted multilayer network, and we do not distinguish the layers. In this case, the edge weights in the aggregated network will be the rate of presence across the layers. For example, if an edge is present in the first three layers, and another edge is present in layers 3, 4, and 5, then we expect the weights of the two edges are the same in the aggregated network.

\begin{figure}[ht]
\includegraphics[width=\linewidth]{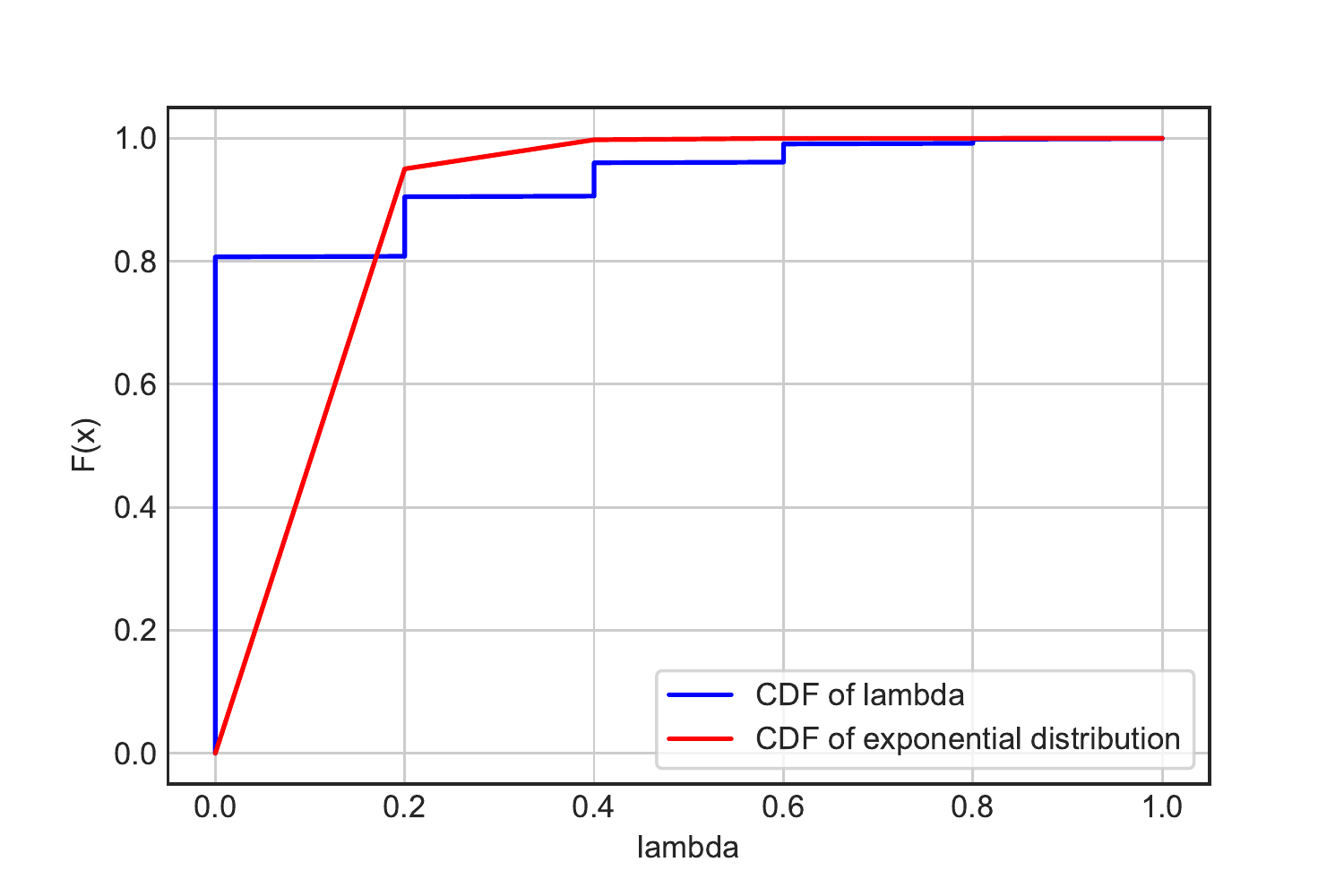}
 \caption{Edge weights of the aggregated network and the corresponding exponential distribution.}
 \label{fig:fig6}
\end{figure}

The network we use to demonstrate the method is a friendship network that consists of 5 layers. The nodes in the multilayer network are the employees of the Department of Computer Science at Aarhus University, Denmark. The five layers describe online and offline relationships between the employees, i.e., Facebook, Leisure, Work, Co-authorship, and Lunch. In Fig. \ref{fig:fig6}, the blue curve shows the CDF of the distribution of edge weights. The CDF of the corresponding exponential distribution is the red curve in Fig. \ref{fig:fig6}. As we can see, there are five steps, which corresponding to the rate of presence of edges in the five layers.

\section{Conclusion}
In this paper, we proposed a \emph{maximum a posteriori} estimation-based algorithm to aggregate mutlilayer networks. In constructing single-layer network from various datasets, we first need to obtain a weighted network for each dataset. The edge weights between node pairs are assumed to follow the Poisson distribution across all the layers. Also, the edge weights in the target network are thought to follow the exponential distribution. We apply the expectation-maximization algorithm to find the optimized edge weights and rate parameter of the exponential distribution. 

In the weighted network example, the algorithm converges quickly. The ultimate rate parameter and expected edge weights are affected by the number of datasets. Both the coefficients of determination and rate parameters become stable when there are at least 4 datasets. To evaluate the target network, we adopted the Von Neumann entropy to measure the mixedness of the network. In section \ref{sec3}, we show that the network obtained through our method conserves the core information of the multilayer network. In the unweighted network example, where the edges are not distinguishable, the edge weight in aggregated network is exactly the rate of presence across all of the layers. 

We compared the method to a network obtained through conflated datasets and a network by averaging edge weights directly. The advantage of the proposed method is that we are not required to normalize the expression values, and thus the systemic errors that come from platforms can be reduced. However, compared to the ConNet, the drawback of the proposed method is that there are must be enough conditions in each dataset to find the edge weights in our example. The experiments that contain less than 10 conditions are not taken into consideration, which is the drawback of using a multilayer network.

\ifCLASSOPTIONcompsoc
  \section*{Acknowledgments}
\else
  \section*{Acknowledgment}
\fi
This works was supported by the National Institutes
of Health under Grant No. 1R01AI140760-01A1.

\ifCLASSOPTIONcaptionsoff
  \newpage
\fi

\end{document}